\documentclass[aps,prl,twocolumn,superscriptaddress]{revtex4-1}
\usepackage{graphicx}
\usepackage{amsmath}
\usepackage{psfrag}
\usepackage{float}
\usepackage{subfig}
\newsavebox{\measurebox}
\graphicspath{%
	{converted_graphics/}% inserted by PCTeX
	{/}% inserted by PCTeX
}
\let\oldhat\hat
\newcommand{\vect}[1]{\boldsymbol{#1}}
\renewcommand{\hat}[1]{\oldhat{\boldsymbol{#1}}}
\begin{document}

%\preprint{}

\title{TE-wave propagation in a graded waveguide structure}


\author{Mariana Dalarsson*}
\author{Sven Nordebo}
\affiliation{Department of Physics and Electrical Engineering, Linnaeus University, 351 95 V\"axj\"o, Sweden.}
\email[Corresponding author e-mail: ]{mariana.dalarsson@lnu.se}
%\author{Zoran Jak\v{s}i\'{c}}
%\affiliation{Center of Microelectronic Technologies and Single Crystals, Institute of Chemistry, Technology and Metallurgy, University of Belgrade, 11 000 Belgrade, Serbia}
%\thanks{}


\date{\today}

\begin{abstract}
We investigate TE-wave propagation in a hollow waveguide with a graded dielectric layer, described using a hyperbolic tangent function. General formulae for the electric field components of the TE-waves, applicable to hollow waveguides with arbitrary cross sectional shapes, are presented. We illustrate the exact analytical results for the electric field components in the special case of a rectangular waveguide. Furthermore, we derive exact analytical results for the reflection and transmission coefficients valid for waveguides of arbitrary cross sectional shapes. Finally, we show that the obtained reflection and transmission coefficients are in exact asymptotic agreement with those obtained for a very thin homogeneous dielectric layer using mode-matching and cascading. The proposed method is tractable since it gives analytical results that are directly applicable without the need of mode-matching, and it has the ability to model realistic, smooth transitions.  
\end{abstract}

\pacs{41.20.-q}
%\keywords{}

\maketitle

\section{Introduction}
Recent studies of radio frequency absorption and optimal plasmonic resonances in gold nanoparticle (GNP) suspensions \cite{dalarsson1,ivanenko1,nordebo} have given rise to an interest in plasmonic resonances in layered waveguide structures. In particular, the scattering on a single thin layer, modeled as a thin dielectric layer in a straight waveguide, with perfectly electrically conducting (PEC) boundaries and a homogeneous cross section with material parameters $\epsilon$ and $\mu$, is reported in \cite{ivanenko2}. Following a number of previous studies by one of the present authors \cite{dalarsson2,dalarsson3,dalarsson4,dalarsson5,dalarsson6,dalarsson7,dalarsson8}, in this paper the surrounding homogeneous straight waveguide medium with a single thin layer is described as a stratified medium with frequency-dependent permittivity $\epsilon = \epsilon(\omega, z)$ being a function of the waveguide axis direction (chosen to be the $z$-direction). One important feature of the present approach to the TE-wave scattering on a thin dielectric layer in a hollow waveguide, is that it is possible to obtain the total scattering matrix parameters in the entire waveguide structure without any need to use boundary conditions, mode matching and cascading techniques. The waveguide is treated as filled with a single composite material with stratified frequency-dependent permittivity. Thus, a single solution of Maxwell's equations in such a material replaces partial solutions in different materials, while at the same time asymptotically approaching such partial solutions in different materials. Furthermore, the boundary conditions between materials are built in into the stratified permittivity function, and are hence not needed.     

Regarding notation and conventions, we consider classical electrodynamics where the electric and magnetic fields $\vect{E}$ and 
$\vect{H}$, respectively, are given in SI-units. The time convention for time harmonic fields (phasors) is given by $\exp(\mathrm{j} \omega t)$ where $\omega$ is the angular frequency and $t$ the time. We assume time-harmonic fields in a non-magnetic ($\mu = \mu_0 \mu_R$ with 
$\mu_R = 1$) inhomogeneous isotropic waveguide material.

\section{Problem formulation}
\begin{figure}[t] 
	\centering
	\includegraphics[scale=0.45, trim={50 250 50 50}, clip]{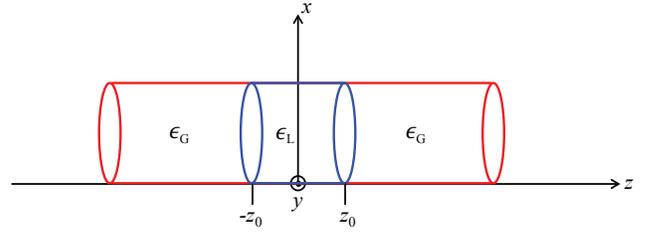}
	\caption{Hollow waveguide with a dielectric layer}
	\label{Figure1}
\end{figure}

\begin{figure}[h]
	\centering
	\includegraphics[scale=0.41]{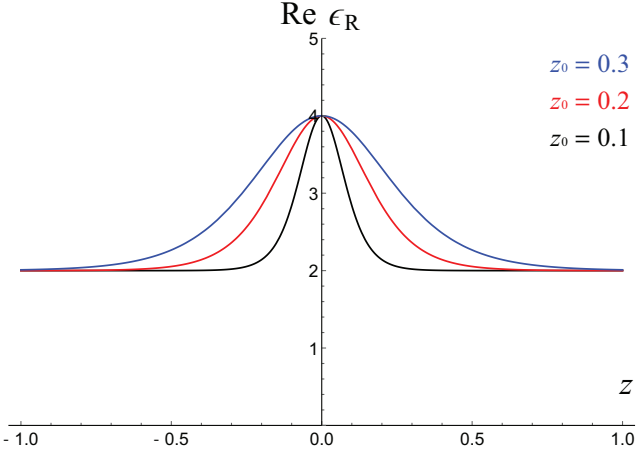}
	\caption{Three examples of permittivity functions changing from $\Re[\epsilon_\mathrm{G}] = 2$ to $\Re[\epsilon_\mathrm{L}] = 4$ and back for $z_0 = 0.1$ (black line), $z_0 = 0.2$ (red line) and $z_0 = 0.3$ (blue line). Here $\Re[\epsilon_\mathrm{L}(\omega)] > \Re[\epsilon_\mathrm{G}(\omega)]$. Note, however, that this assumption is not essential for the present approach, and is only used for graphical illustration.}
	\label{Figure2}
\end{figure}

The geometry of the problem is illustrated in Fig.~\ref{Figure1}. In the surrounding non-magnetic lossy homogeneous straight waveguide medium with complex relative permittivity $\epsilon_\mathrm{G}(\omega)$, a single lossy non-magnetic thin layer with complex relative permittivity 
$\epsilon_\mathrm{L}(\omega)$ is inserted about the plane $z$ = 0, as shown in 
Fig.~\ref{Figure1}. The proposed model is applicable to any complex permittivities of the two media, including negative values in chiral metamaterials, as long as they satisfy the Kramers-Kronig relations. Mathematically, the waveguide medium can be described as a stratified medium with frequency-dependent permittivity $\epsilon = \epsilon(\omega, z)$ given by the following function of the waveguide axis direction (chosen to be the $z$-direction), 
\begin{eqnarray*}
\epsilon(\omega, z) = \epsilon_0 \epsilon_\mathrm{R}(z)=
\end{eqnarray*}
\begin{equation}
\epsilon_0 \left\{ \epsilon_\mathrm{L}(\omega) - \left[\epsilon_\mathrm{L}(\omega) - \epsilon_\mathrm{G}(\omega) \right] 
\tanh^2 \left( \frac{z}{z_0} \right) \right\} \hspace{1mm} ,
\label{Eq1}
\end{equation} 
where $\epsilon_\mathrm{R}(z)$ denotes the relative permittivity, and $2 z_0$ determines the size of the inserted layer about the plane $z$ = 0, as indicated in Fig.~\ref{Figure1}. 
Far away from the layer in both directions ($z \to \pm \infty$), we have  
\begin{equation}
\tanh^2 \left( \frac{z}{z_0} \right) \to 1 \hspace{1mm} \Rightarrow \hspace{1mm} \epsilon(\omega, \pm \infty) = \epsilon_0 \epsilon_\mathrm{G}(\omega) \hspace{1mm} ,
\label{Eq2}
\end{equation} 
while at the layer ($z \to 0$), we have
\begin{equation}
\tanh^2 \left( \frac{z}{z_0} \right) \to 0 \hspace{1mm} \Rightarrow \hspace{1mm} \epsilon(\omega, 0) = \epsilon_0 \epsilon_\mathrm{L}(\omega) \hspace{1mm} ,
\label{Eq3}
\end{equation} 
as required by the geometry of the problem. A geometry with a very thin single layer with rapid smooth transition from $\epsilon_\mathrm{G}(\omega)$ to $\epsilon_\mathrm{L}(\omega)$ and back to $\epsilon_\mathrm{G}(\omega)$ is then obtained in the limit $z_0 \to 0$. A few examples of permittivity functions for different values of $z_0$, are shown in Fig.~\ref{Figure2}.

\begin{figure}[h] 
	\centering
	\includegraphics[scale=0.8, trim={160 140 50 260}, clip]{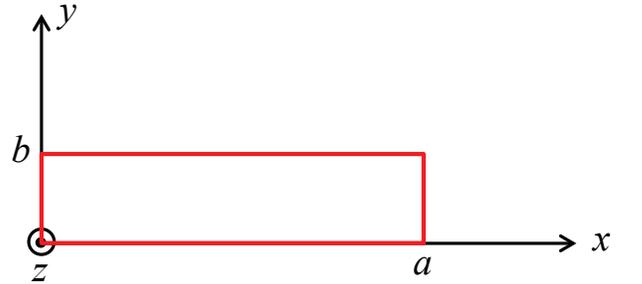}
	\caption{Cross section of a rectangular waveguide with dimensions $a$ and $b$ such that $a > b$.}
	\label{Figure3}
\end{figure}

Wave propagation in a waveguide, with no field sources inside ($\rho = 0$ , $\vect{J} = 0$), is governed by Maxwell's equations
\begin{eqnarray*}
\nabla \times \vect{E} = - \textrm{j} \omega \mu_0 \vect{H} \hspace{2mm} , \hspace{2mm} \nabla \cdot \left[ \epsilon(z) \vect{E} \right] = 0 \hspace{1mm} ,
\end{eqnarray*}
 \begin{equation}
\nabla \times \vect{H} = \textrm{j} \omega \epsilon(z) \vect{E} \hspace{2mm} , \hspace{2mm} \nabla \cdot \vect{H} = 0 \hspace{1mm} .
\label{Eq4}
\end{equation}
The Maxwell equations~(\ref{Eq4}) give rise to the following wave equations for the electric and magnetic fields $\vect{E}$ and $\vect{H}$, respectively,
\begin{equation}
\nabla^2 \vect{E} + \nabla \left( \frac{1}{\epsilon_\mathrm{R}} \frac{\mathrm{d} \epsilon_\mathrm{R}}{\mathrm{d}z} E_z \right) + k^2 \epsilon_\mathrm{R}(z) \vect{E} = 0 \hspace{1mm} ,
\label{Eq5}
\end{equation}
\begin{equation}
\nabla^2 \vect{H} + \frac{1}{\epsilon_\mathrm{R}} \frac{\mathrm{d} \epsilon_\mathrm{R}}{\mathrm{d}z} \left( \nabla H_z - \frac{\partial \vect{H}}{\partial z} \right) + k^2 \epsilon_\mathrm{R}(z) \vect{H} = 0 \hspace{1mm} ,
\label{Eq6}
\end{equation}
where $k^2 = \omega^2 \epsilon_0 \mu_0 = \omega^2 / c^2$. For TE-waves with $E_z = 0$, the wave equations~(\ref{Eq5})-(\ref{Eq6}) become simply
\begin{equation}
\nabla^2 \vect{E} + k^2 \epsilon_\mathrm{R}(z) \vect{E} = 0 \hspace{1mm} ,
\nabla^2 H_z + k^2 \epsilon_\mathrm{R}(z) H_z = 0 \hspace{1mm}.
\label{Eq7}
\end{equation}
It is possible to solve the second of the equations~(\ref{Eq7}) for the longitudinal component of the magnetic field $H_z$, and from that solution obtain all the other field components using standard waveguide analysis techniques. On the other hand, it is also possible to solve the first of the equations ~(\ref{Eq7}) for the electric field $\vect{E}$, whereby the magnetic field $\vect{H}$ is readily obtained from the first of Maxwell's equations~(\ref{Eq4}), i.e. using        
\begin{equation}
\vect{H} = \frac{\textrm{j}}{\omega \mu_0} \nabla \times \vect{E} \hspace{1mm} .
\label{Eq8}
\end{equation}
Thus, the first of the equations~(\ref{Eq7}) with $E_z = 0$ is equivalent to two scalar equations, both of which are of the form
\begin{equation}
\nabla^2 E_j + k^2 \epsilon_\mathrm{R}(z) E_j = 0 \hspace{1mm} , \hspace{1mm} j \in \{ x , y \} \hspace{1mm} .
\label{Eq9}
\end{equation}
By means of standard separation of variables $E_j = F_j(x,y) Z(z)$, any of these two differential equations can be split into an equation for $F_j(x,y)$ and an equation for $Z(z)$ 
\begin{equation}
\left( \frac{\partial^2}{\partial x^2} + \frac{\partial^2}{\partial y^2} \right) F_j + k_T^2 F_j = 0 \hspace{1mm} , \hspace{1mm} j \in \{ x , y \} \hspace{1mm} ,
\label{Eq10}
\end{equation}
\begin{equation}
\frac{\mathrm{d}^2 Z}{\mathrm{d}z^2} + (k^2 \epsilon_\mathrm{R}(z) - k_T^2) Z = 0 \hspace{1mm} ,
\label{Eq11}
\end{equation}
where $k_T$ denotes the transverse wave number of the waveguide. The solutions of the equations~(\ref{Eq10}) are the standard solutions obtained for a particular waveguide cross sectional shape, and are unaffected by the graded material transition in the $z$-direction. Let us, as an example, consider the rectangular waveguide, with a cross section shown in Fig.~\ref{Figure3}.    
%\begin{figure}[tbp] 
%  \centering
%  \includegraphics[width=3.94in,height=1.99in,keepaspectratio]{Figure3}
%  \caption{Cross section of a rectangular waveguide with dimensions $a$ and $b$ such that $a > b$.}
%  \label{Figure3}
%\end{figure}
\\
Equation~(\ref{Eq10}) then has the well-known solutions for a rectangular waveguide
\begin{equation}
F_x = A \left( \frac{n \pi}{b} \right) \cos \left( \frac{m \pi x}{a} \right) \sin \left( \frac{n \pi y}{b} \right) \hspace{1mm} ,
\label{Eq12}
\end{equation} 
\begin{equation}
F_y = - A \left( \frac{m \pi}{a} \right) \sin \left( \frac{m \pi x}{a} \right) \cos \left( \frac{n \pi y}{b} \right) \hspace{1mm} ,
\label{Eq13}
\end{equation}
where the well-known result for the square of the transverse wave number is
\begin{equation}
k_T^2 = \left( \frac{m \pi}{a} \right)^2 + \left( \frac{n \pi}{b} \right)^2 \hspace{1mm} ,
\label{Eq14}
\end{equation}
and the constant $A$ is to be determined from the incoming field intensity $E_0 = E(- \infty)$. The equations~(\ref{Eq10}) are quite general, and upon choosing any other hollow waveguide (e.g. parallel-plate waveguide, coaxial waveguide etc.) we can simply reuse the existing results for the transverse functions, whenever they are available in closed form. The waves can propagate in the waveguide only if 
\begin{equation}
k^2 \Re\{\epsilon_\mathrm{R}(z) \} - k_T^2 > 0 \hspace{1mm} \Rightarrow \hspace{1mm}  k^2 [\Re\{\epsilon_\mathrm{R}(z) \}]_\text{min} - k_T^2 > 0 \hspace{1mm} .
\label{Eq15}
\end{equation}
In the particular case where we have assumed $\Re[\epsilon_\mathrm{L}(\omega)] > \Re[\epsilon_\mathrm{G}(\omega)] > 0$, (\ref{Eq15}) yields the following condition 
\begin{equation}
k^2 \Re\{\epsilon_\mathrm{G} \} - k_T^2 > 0 \hspace{0.7mm} \Rightarrow \hspace{0.7mm}  k^2 \Re\{\epsilon_\mathrm{G}\} > k_T^2 \hspace{0.7mm} \Rightarrow \hspace{0.7mm} \omega^2 > \frac{k_T^2 c^2}{\Re\{\epsilon_\mathrm{G} \}} \hspace{0.7mm} ,
\label{Eq16}
\end{equation}
or
\begin{equation}
f > \frac{c}{2 \pi} \frac{k_T}{\sqrt{\Re\{\epsilon_\mathrm{G} \}}} = f_{c,\text{max}} \hspace{1mm} .
\label{Eq17}
\end{equation}
Thus, with our assumptions, the waves can only propagate if their frequency is higher than the maximum cutoff frequency $f_{c,\text{max}}$ defined in equation~(\ref{Eq17}) above. It is interesting to note that in the stratified media model employed here, the cutoff frequency is a function of the spatial $z$-coordinate, i.e. we can write $f_c(z)=c/(2 \pi) \cdot k_T/\sqrt{\Re\{\epsilon_\mathrm{R}(z)\}}$. Furthermore, with our choice of permittivities ($\Re[\epsilon_\mathrm{L}(\omega)] > \Re[\epsilon_\mathrm{G}(\omega)]$ in the entire operating frequency range), waves with lower frequencies than $f_{c,\text{max}}$ could propagate in the layer region about $z = 0$. However, such waves could only be trapped in the layer region, and would never be able to propagate either to or from the layer region. Thus, the waves that can propagate through the entire waveguide must necessarily have frequencies higher than $f_{c,\text{max}}$.

\section{Solution of the Longitudinal Equation}

The next objective is to find the solutions of the longitudinal equation~(\ref{Eq11}), which has the form
\begin{equation}
\frac{\mathrm{d}^2 Z}{\mathrm{d}w^2} + \left( D - B \tanh^2 w \right) Z = 0 \hspace{1mm} ,
\label{Eq18}
\end{equation}   
where we introduced a dimensionless variable $w = z/z_0$ and the two dimensionless functions
\begin{equation}
D = (k^2 \epsilon_\mathrm{L} - k_T^2) z_0^2 \hspace{1mm} , \hspace{1mm} B = k^2 z_0^2 (\epsilon_\mathrm{L} - \epsilon_\mathrm{G}) \hspace{1mm} .
\label{Eq19}
\end{equation}  
Using analogous solution procedures to the ones used in our previous work on graded metamaterials \cite{dalarsson2}-\cite{dalarsson8}, we readily obtain  
\begin{eqnarray*}
	Z(z) = T \exp \left( 2 p \hspace{1mm} \frac{z}{z_0} \right) \left[ 1 + \exp \left( 2 \hspace{1mm} \frac{z}{z_0} \right) \right]^{-2p} \cdot
\end{eqnarray*}
\vspace{-3mm}
\begin{eqnarray*}
\cdot_2 F_1 \left[2p + \frac{1}{2} + \sqrt{r^2 + \frac{1}{4}} \hspace{1mm} , \hspace{1mm} 2p + \frac{1}{2} - \sqrt{r^2 + \frac{1}{4}} \hspace{1mm} , \hspace{1mm} 2p + 1 \hspace{1mm} \right.
\end{eqnarray*}
\vspace{-3mm}
\begin{equation}; \hspace{1mm} 
\left. \frac{1}{1 + \exp \left( 2 \hspace{1mm} z/z_0 \right)} \right] ,
\label{Eq20}
\end{equation} 
where $T$ is a constant to be determined from the asymptotic behavior of the solution~(\ref{Eq20}) far away from the layer ($z \to \pm \infty$) 
$\hspace{1mm} \textrm{and}_{\hspace{2mm} 2}F_1(a,b,c;u) = F(a,b,c;u)$ is the ordinary Gaussian hypergeometric function defined by Gauss hypergeometric series \cite{abramowitz}   
\begin{equation}
F(a,b,c;u) = \frac{\Gamma(c)}{\Gamma(a) \Gamma(b)} \sum_{n=0}^{\infty} \frac{\Gamma(a+n) \Gamma(b+n)}{\Gamma(c+n)} \hspace{1mm} \frac{u^n}{n!} \hspace{1mm} ,
\label{Eq21}
\end{equation} 
$\Gamma$ is the Gamma function \cite{abramowitz}, and we define two dimensionless constants $p$ and $r$, as follows
\begin{equation}
p = \textrm{j} \hspace{1mm} \frac{z_0}{2} \hspace{1mm} \sqrt{k^2 \epsilon_\mathrm{G} - k_T^2} = \textrm{j} \hspace{1mm} \frac{k_{z\mathrm{G}} z_0}{2} \hspace{1mm} , \hspace{1mm} 
r = k \hspace{1mm} z_0 \hspace{1mm} \sqrt{\epsilon_\mathrm{L} - \epsilon_\mathrm{G}} \hspace{1mm} ,
\label{Eq22}
\end{equation} 
with $k_{z\mathrm{G}} = \sqrt{k^2 \epsilon_\mathrm{G} - k_T^2}$ being the $z$-component of the wave vector of the asymptotic waves for $z \to \pm \infty$ .
On the other hand, the $z$-component of the wave vector about the origin, where the thin dielectric layer is situated, is denoted by 
$k_{z\mathrm{L}} = \sqrt{k^2 \epsilon_\mathrm{L} - k_T^2}$. Let us now investigate the asymptotic behavior of the solution~(\ref{Eq20}) for 
$z \to + \infty$, when the argument of the hypergeometric function becomes zero, i.e.
\begin{equation}
u = \frac{1}{1 + \exp \left( 2 \hspace{1mm} z/z_0 \right)} \to 0 \hspace{1mm} \textrm{for} \hspace{1mm} z \to + \infty \hspace{1mm} .
\label{Eq23}
\end{equation} \\
Using the series~(\ref{Eq21}), we see that $F(a,b,c;0) = 1$, and we obtain from~(\ref{Eq20}), \\
\begin{equation} 
Z(z) \to T \exp \left( - \textrm{j} \hspace{1mm} k_{z\mathrm{G}} \hspace{1mm} z \right) \hspace{1mm} \textrm{for} \hspace{1mm} z \to + \infty \hspace{1mm} ,  
\label{Eq24}
\end{equation} 
being a transmitted forward-propagating wave with amplitude equal to one, as required. Next, we investigate the asymptotic behavior of 
the solution~(\ref{Eq20}) for $z \to - \infty$, when the argument of the hypergeometric function becomes equal to one, i.e.
\begin{equation}
u = \frac{1}{1 + \exp \left( 2 \hspace{1mm} z/z_0 \right)} \to 1 \hspace{1mm} \textrm{for} \hspace{1mm} z \to - \infty \hspace{1mm} .
\label{Eq25}
\end{equation} \\
In order to investigate the asymptotic behavior of the solution~(\ref{Eq20}) for $z \to - \infty$, it is convenient to use the following transformation formula for hypergeometric functions \cite{abramowitz}
\begin{eqnarray*}
	F(a,b,c;u) \hspace{-0.5mm}=\hspace{-0.5mm} \frac{\Gamma(c) \Gamma(c-a-b)}{\Gamma(c-a) \Gamma(c-b)} F(a,b,a+b-c+\hspace{-0.2mm}1;\hspace{-0.5mm}1-u) 
\end{eqnarray*}
\begin{eqnarray*}
	+ (1-u)^{c-a-b} \hspace{0.5mm}\cdot
\end{eqnarray*}
\begin{equation}
\frac{\Gamma(c) \Gamma(a+b-c)}{\Gamma(a) \Gamma(b)} F(c-a,c-b,c-a-b+1;1-u)  ,
\label{Eq26}
\end{equation} 
such that in the limit $z \to - \infty$, with $c-a-b = - 2p$, we obtain 
\begin{eqnarray*}
Z(z) \to T \frac{\Gamma(c) \Gamma(c-a-b)}{\Gamma(c-a) \Gamma(c-b)} \exp \left( + \textrm{j} \hspace{1mm} k_{z\mathrm{G}} \hspace{1mm} z \right) 
\end{eqnarray*}
\begin{equation} 
+ 
T \frac{\Gamma(c) \Gamma(a+b-c)}{\Gamma(a) \Gamma(b)} \exp \left( - \textrm{j} \hspace{1mm} k_{z\mathrm{G}} \hspace{1mm} z \right)
\hspace{1mm} \textrm{for} \hspace{1mm} z \to - \infty \hspace{1mm} .  
\label{Eq27}
\end{equation} 
where we require the solution to be a combination of an incident TE-wave and a reflected TE-wave with amplitude equal to one, as follows
\begin{equation} 
Z(z) \to \exp \left( - \textrm{j} \hspace{1mm} k_{z\mathrm{G}} \hspace{1mm} z \right) + R \exp \left( + \textrm{j} \hspace{1mm} k_{z\mathrm{G}} \hspace{1mm} z \right) 
\hspace{0.7mm} \textrm{for} \hspace{0.8mm} z \to - \infty \hspace{0.7mm} ,  
\label{Eq28}
\end{equation} 
with the notation
\begin{equation} 
a = 2p + \frac{1}{2} + \sqrt{r^2 + \frac{1}{4}} \hspace{0.7mm} , \hspace{0.7mm} b = 2p + \frac{1}{2} - \sqrt{r^2 + \frac{1}{4}} \hspace{0.7mm} , \hspace{0.7mm} c = 2p +1 \hspace{0.7mm} .
\label{Eq29}
\end{equation} 
Comparing the equations~(\ref{Eq27}) and~(\ref{Eq28}), we readily obtain the general expressions for the transmission coefficient ($T$) and the reflection coefficient ($R$), in the form    
\begin{eqnarray*}
T = \frac{\Gamma(a) \Gamma(b)}{\Gamma(c) \Gamma(a+b-c)} \hspace{1mm} ,
\end{eqnarray*}
\begin{equation} 
R = \frac{\Gamma(a) \Gamma(b)}{\Gamma(c-a) \Gamma(c-b)} \frac{\Gamma(c-a-b)}{\Gamma(a+b-c)} \hspace{1mm} .
\label{Eq30}
\end{equation} 
The results~(\ref{Eq30}) are the most general exact analytic results for transmission and reflection coefficients over a graded dielectric layer in a straight hollow waveguide, valid for waveguides with any cross sectional shape. In the special case of a rectangular waveguide, we then obtain the overall expressions for the two electric field components in the form     
\begin{eqnarray*}
	E_x = A \hspace{1mm} T \left( \frac{n \pi}{b} \right) \cos \left( \frac{m \pi x}{a} \right) \sin \left( \frac{n \pi y}{b} \right) \exp \left( 2 p \hspace{1mm} \frac{z}{z_0} \right) \cdot
\end{eqnarray*}
\vspace{-4mm}
\begin{eqnarray*}
	\left[ 1 + \exp \left( 2 \hspace{1mm} \frac{z}{z_0} \right) \right]^{-2p} \cdot
\end{eqnarray*}
\vspace{-4mm}
\begin{eqnarray*}
\cdot_2 F_1 \left[2p + \frac{1}{2} + \sqrt{r^2 + \frac{1}{4}} \hspace{1mm} , \hspace{1mm} 2p + \frac{1}{2} - \sqrt{r^2 + \frac{1}{4}} \hspace{1mm} , \hspace{1mm} 2p + 1 \hspace{1mm} \right.	
\end{eqnarray*}
\vspace{-4mm}
\begin{equation}; \hspace{1mm} \left.
\frac{1}{1 + \exp \left( 2 \hspace{1mm} z/z_0 \right)} \right] ,
\label{Eq31}
\end{equation} 
\begin{eqnarray*}
	E_y = - A \hspace{1mm} T \left( \frac{m \pi}{a} \right) \sin \left( \frac{m \pi x}{a} \right) \cos \left( \frac{n \pi y}{b} \right) \exp \left( 2 p \hspace{1mm} \frac{z}{z_0} \right) 
\end{eqnarray*}
\vspace{-4mm}
	\begin{eqnarray*}
	\left[ 1 + \exp \left( 2 \hspace{1mm} \frac{z}{z_0} \right) \right]^{-2p} \cdot
\end{eqnarray*}
\vspace{-4mm}
\begin{eqnarray*}
\cdot_2 F_1 \left[2p + \frac{1}{2} + \sqrt{r^2 + \frac{1}{4}} \hspace{1mm} , \hspace{1mm} 2p + \frac{1}{2} - \sqrt{r^2 + \frac{1}{4}} \hspace{1mm} , \hspace{1mm} 2p + 1 \hspace{1mm} \right.
\end{eqnarray*}
\vspace{-4mm}
\begin{equation}
; \hspace{1mm} \left.
\frac{1}{1 + \exp \left( 2 \hspace{1mm} z/z_0 \right)} \right] ,
\label{Eq32}
\end{equation} 
where $A$ is a constant proportional to the incident electric field amplitude $E_0$. The magnetic field components in a rectangular waveguide are then readily obtained using the Maxwell equation~(\ref{Eq8}). 

In the present paper, we have studied TE-wave propagation, since the Helmholtz equations for the longitudinal component of the magnetic field and the transverse components of the electric field (\ref{Eq7}) allow for exact analytical solutions in the case of non-magnetic stratified media with frequency-dependent permittivity $\epsilon = \epsilon(\omega, z)$. Although TM-wave propagation is also of interest for the investigations reported in \cite{ivanenko2}, the Helmholtz equations (\ref{Eq5})-(\ref{Eq6}) for TM-wave propagation are mathematically more complex, and their analytical solutions require somewhat different solution techniques. The analytical solutions for the case of TM-wave propagation will therefore be the subject of a future publication. 

\section{Asymptotic analysis}
It is now of interest to study the transmission and reflection coefficients~(\ref{Eq30}) in the case of a thin dielectric layer 
($z_0 \to 0$), when both constants $p$ and $r$ approach zero. Using the properties of the Gamma function \cite{abramowitz} and 
(\ref{Eq29}), with the assumption $z_0 \to 0$, we obtain from the results~(\ref{Eq30})
\begin{equation}
T = 1 + \textrm{j} k_{z\mathrm{L}} z_0 \hspace{1mm} 
\frac{\epsilon_\mathrm{L} - \epsilon_\mathrm{G}}{\sqrt{\epsilon_\mathrm{L} - k_T^2/k^2} \sqrt{\epsilon_\mathrm{G} - k_T^2/k^2}} + {\cal O}\{z_0^2\}\hspace{1mm} ,
\label{Eq33}
\end{equation}
\begin{equation}
R = \textrm{j} k_{z\mathrm{L}} z_0 \hspace{1mm} \frac{\epsilon_\mathrm{L} - \epsilon_\mathrm{G}}{\sqrt{\epsilon_\mathrm{L} - k_T^2/k^2} \sqrt{\epsilon_\mathrm{G} - k_T^2/k^2}} +{\cal O}\{z_0^2\}
\hspace{1mm} ,
\label{Eq34}
\end{equation}
where $T=1+R$ as required. From the results~(\ref{Eq33}-\ref{Eq34}), we readily see that there is no reflection ($R = 0$) whenever the two materials have the same relative permittivity ($\epsilon_\mathrm{L}(\omega) = \epsilon_\mathrm{G}(\omega)$), e.g. the two materials are the same. Furthermore, we see that for 
$z_0 = 0$, which implies that the dielectric layer is removed, there is no reflection either ($R = 0$), as expected.

Finally, it is of interest to compare the asymptotic result~(\ref{Eq34}) with the corresponding result obtained in \cite{ivanenko2} for a non-graded layered waveguide structure using mode-matching and cascading methods for hollow waveguides. The reflection coefficient reported in \cite{ivanenko2} is denoted by $T_{11}^{(2)}$, and using the notation employed in the present paper, has the form
\begin{equation}
T_{11}^{(2)} = - 2 \textrm{j} k_{z\mathrm{L}} (2 z_0) \frac{S_{11}^{(1)}}{1 - \left( S_{11}^{(1)} \right)^2} +{\cal O}\{z_0^2\}\hspace{1mm} ,
\label{Eq35}
\end{equation}
where $2 z_0 = d_\mathrm{L}$ is the thickness of the thin dielectric layer, $\mu_\mathrm{G} = \mu_\mathrm{L} = 1$ and 
\begin{equation}
S_{11}^{(1)} = \frac{k_{z\mathrm{G}} - k_{z\mathrm{L}}}{k_{z\mathrm{G}} + k_{z\mathrm{L}}} \hspace{1mm} .
\label{Eq36}
\end{equation}
Substituting~(\ref{Eq36}) into the result~(\ref{Eq35}), after some algebra, we obtain
\begin{equation}
T_{11}^{(2)} = - 2 \hspace{1mm} \textrm{j} k_{z\mathrm{L}} (2 z_0) \frac{k_{z\mathrm{G}}^2 - k_{z\mathrm{L}}^2}{4 \hspace{1mm} k_{z\mathrm{G}} k_{z\mathrm{L}}} 
+{\cal O}\{z_0^2\}\hspace{1mm} ,
\label{Eq37}
\end{equation}
Using the definitions of $k_{z\mathrm{G}}$ and $k_{z\mathrm{L}}$ stated earlier in this paper, and inserting them into equation~(\ref{Eq35}), we obtain 
\begin{equation}
T_{11}^{(2)} = R = \textrm{j} k_{z\mathrm{L}} z_0 \hspace{0.5mm} 
\frac{\epsilon_\mathrm{L} - \epsilon_\mathrm{G}}{\sqrt{\epsilon_\mathrm{L} - k_T^2/k^2} \sqrt{\epsilon_\mathrm{G} - k_T^2/k^2}} +{\cal O}\{z_0^2\}\hspace{0.7mm} .
\label{Eq38}
\end{equation}
From the results~(\ref{Eq38}) and~(\ref{Eq34}), we see that the scattering matrix parameters reported in \cite{ivanenko2} for a homogeneous dielectric layer in a hollow waveguide structure, have the same thin layer asymptotics as the scattering parameters obtained here using the graded dielectric layer based on~(\ref{Eq1}). This concludes the asymptotic analysis, and illustrates that the graded permittivity function~(\ref{Eq1}) can be employed  to obtain useful scattering parameters for the waveguide without any need of mode matching and cascading techniques. Furthermore, the proposed technique gives the flexibility to model realistic, smooth transitions. 

\section{Conclusions}
We investigated TE-wave propagation in a hollow waveguide with a graded dielectric layer, described using a hyperbolic tangent function. General formulae for the electric field components of TE-waves, applicable to hollow waveguides with arbitrary cross sections, were obtained. Furthermore, we obtained exact analytical results for the electric field components in the special case of the rectangular waveguide, as well as the exact analytical results for reflection and transmission coefficients valid for waveguides with arbitrary cross sectional shapes. Finally, we showed that the obtained reflection and transmission coefficients are in exact asymptotic agreement with those obtained in \cite{ivanenko2} for a very thin homogeneous dielectric layer using mode-matching and cascading. The proposed method is tractable since it gives analytical results that are directly applicable without the need of mode-matching. At the same time, our method has the ability to model realistic, smooth transitions.

\begin{acknowledgments}
	The work of S. N. was supported by the Swedish Foundation for Strategic Research (SSF) under the program Applied
	Mathematics and the project ``Complex analysis and convex optimization for EM design''.
\end{acknowledgments}

\bibliography{bibliography}

\end{document}